\documentclass[preprint]{aastex}

\newcommand{\ctwohtwo}{\mbox{C$_{2}$H$_{2}$}}
\newcommand{\htwo}{\mbox{\ion{H}{2}}}
\newcommand{\HII}{H\,{\sc ii}}
\newcommand{\hcn}{\mbox{HCN}}
\newcommand{\mgs}{\mbox{MgS}}
\newcommand{\neii}{\mbox{[\ion{Ne}{2}]\,12.8\,\micron}}
\newcommand{\neiiia}{\mbox{[\ion{Ne}{3}]\,15.5\,\micron}}
\newcommand{\neiiib}{\mbox{[\ion{Ne}{3}]\,36.0\,\micron}}
\newcommand{\ariii}{\mbox{[\ion{Ar}{3}]\,9.0\,\micron}}
\newcommand{\siv}{\mbox{[\ion{S}{4}]\,10.5\,\micron}}
\newcommand{\siiia}{\mbox{[\ion{S}{3}]\,18.7\,\micron}}
\newcommand{\siiib}{\mbox{[\ion{S}{3}]\,33.5\,\micron}}
\newcommand{\silii}{\mbox{[\ion{Si}{2}]\,34.8\,\micron}}
\newcommand{\lsun}{\mbox{L$_{\odot}$}}

\newcommand{\lir}{\mbox{L$_{\rm IR}$}}

\newcommand{\egant}{\mbox{EVP01}}
\newcommand{\eganp}{\mbox{(EVP01)}}
\newcommand{\spitzer}{\emph{Spitzer}}

\newcommand{\buc}{\mbox{Paper~{\sc I}}}
\newcommand{\kas}{\mbox{Paper~{\sc II}}}

\shorttitle{\emph{Spitzer} IRS spectra of LMC sources}
\shortauthors{Buchanan et al.}

\begin{document}

\title{\emph{Spitzer} IRS Spectra of Luminous 8~\micron\ Sources in
  the Large Magellanic Cloud: Testing color-based classifications}

\author{Catherine L.\ Buchanan\altaffilmark{1}, Joel
  H.\ Kastner\altaffilmark{2}, Bruce J.\ Hrivnak\altaffilmark{3},
  Raghvendra Sahai\altaffilmark{4}}

\altaffiltext{1}{School of Physics, University of Melbourne, 
Parkville, Victoria, 3010 Australia. Email: clb@unimelb.edu.au}
\altaffiltext{2}{Center for Imaging Science, Rochester Institute of
Technology, 54 Lomb Memorial Drive, Rochester NY 14623. Presently Visiting Astronomer, LAOG, Grenoble.}
\altaffiltext{3}{Dept. of Physics and Astronomy, Valparaiso
University, Valparaiso, IN 46383}
\altaffiltext{4}{NASA/JPL, 4800 Oak Grove Drive, Pasadena, CA 91109}

\begin{abstract}
  We present archival \spitzer\ IRS spectra of 19 luminous
  8~\micron\ selected sources in the Large Magellanic Cloud (LMC). The
  object classes derived from these spectra and from an additional 24
  spectra in the literature are compared with classifications based on
  2MASS/MSX (J, H, K, and 8 $\mu$m) colors in order to test the
  ``JHK8'' classification scheme \citep{kas08}. The IRS spectra
  confirm the classifications of 22 of the 31 sources that can be
  classified under the JHK8 system. The spectroscopic classification
  of 12 objects that were unclassifiable in the JHK8 scheme allow us
  to characterize regions of the color-color diagrams that previously
  lacked spectroscopic verification, enabling refinements to the JHK8
  classification system. The results of these new classifications are
  consistent with previous results concerning the identification of
  the most infrared-luminous objects in the LMC. In particular, while
  the IRS spectra reveal several new examples of asymptotic giant
  branch (AGB) stars with O-rich envelopes, such objects are still far
  outnumbered by carbon stars (C-rich AGB stars). We show that
  \protect\spitzer\ IRAC/MIPS color-color diagrams provide improved
  discrimination between red supergiants and oxygen-rich and
  carbon-rich asymptotic giant branch stars relative to those based on
  2MASS/MSX colors. These diagrams will enable the most luminous IR
  sources in Local Group galaxies to be classified with high
  confidence based on their \protect\spitzer\ colors. Such
  characterizations of stellar populations will continue to be
  possible during \protect\spitzer's warm mission, through the use of
  IRAC [3.6]-[4.5] and 2MASS colors.
\end{abstract}


\section{INTRODUCTION} \label{sec:intro}

The infrared wavelength regime is key to understanding the early and
late stages of stellar evolution, which are characterized by
circumstellar dust. Objects in these phases should dominate the
mid-infrared (MIR) point source populations of nearby galaxies.  The
\spitzer\ Space Telescope has enabled the detection of large numbers
of mass-losing evolved stars and dust-enshrouded young stellar objects
throughout the Local Group (e.g., \citealt{blu06, jac06, jac07,
  can06}). Identification of these objects relies on their location in
color-color and color-magnitude diagrams, and is often based on
stellar evolution models (e.g., \citealt{blu06}). Reliable
classification of objects based on photometry requires spectroscopic
confirmation. Such confirmation is especially important when
attempting to distinguish between the various classes of very highly
obscured (and hence IR-luminous) mass-losing evolved stars, such as
red supergiants vs.\ asymptotic giant branch (AGB) stars, and AGB
stars with oxygen-rich vs.\ carbon-rich circumstellar envelopes. These
determinations in turn provide key input for models of galactic
chemical enrichment.

This paper is the third in our series discussing the classification of
compact IR-luminous sources in the Large Magellanic Cloud (LMC).  Our
sample was selected from the sample of $>$1650 2MASS/MSX LMC sources
compiled by \citet{ega01} (hereafter \egant), and comprises the 250
luminous 8~\micron\ sources (F$_{8.3~\micron} > 150$~mJy) that are not
already known to be main sequence stars or Galactic objects
(\citealt{kas08}; hereafter \kas). In \citet[hereafter \buc]{buc06},
we classified and analyzed the \spitzer\ IRS\footnote{The IRS was a
  collaborative venture between Cornell University and Ball Aerospace
  Corporation funded by NASA through the Jet Propulsion Laboratory and
  Ames Research Center.}  \citep{hou04} spectra of 60 sources in our
sample and derived spectroscopically verified color-color diagnostics
using 2MASS and MSX A-band photometry. We refer to
this 2MASS/MSX color-color classification system as the JHK8 scheme.
In \kas\ we classified $>$70\% of the sample with high
confidence, based on the JHK8 color-based classifications supplemented
by data from the literature. In this paper, we use additional IRS
spectra available from the \spitzer\ archive to further refine the
JHK8 scheme and discuss classifications based on \spitzer\ photometry.

Of the sample of 250 luminous 8~\micron\ sources in the LMC, IRS
spectra are available for 123 objects. Classifications for 60 of these
were presented in \buc. This paper presents classifications for 44 of
the remaining 63 objects, which are listed in Table
\ref{tab:sample}. The final 19 archival sources, which were not yet
public at the time of submission of this paper, will be considered in
a future paper.  In \S\ref{sec:obs}, we discuss the data reduction for
20 of the 44 sources. The remaining 24 sources have spectra that are
previously published in \citeauthor{spe06}
\citeyearpar{spe06}, \citeauthor{zij06} \citeyearpar{zij06}, and
\citeauthor{slo08} \citeyearpar{slo08}. We use the published spectra
to do our own independent classification.  In \S\ref{sec:res}, we
provide spectral classifications for these 44 sources (Table
\ref{tab:class}), which break down into the following categories and
subcategories: 17 O-rich objects (3 AGB stars, 3 OH/IR stars, 4 B[e]
stars, and 7 RSGs), 19 C-rich objects (all AGB stars), 4
\htwo\ regions, 3 Young Stellar Objects (YSOs) or candidate YSOs, and
1 unclassified source (spectrum too weak). We present the spectra for
9 new O-rich sources, 4 new C-rich sources, 4 new \htwo\ regions, and
2 candidate YSOs. In \S\ref{sec:dis}, we use these 44 sources to test
and refine the JHK8 scheme. We find that 22 were classified correctly,
9 were not classified correctly, 12 could not be classified with the
JHK8 scheme, and 1 source had a spectrum too weak to classify. We
describe revisions to this classification scheme based on these
results, and extend the classifications to \spitzer\ IRAC/MIPS colors.
We conclude with a summary in \S\ref{sec:con}.

\section{SAMPLE AND DATA REDUCTION} \label{sec:obs}

The selection of the sample is described briefly in \S\ref{sec:intro}
and more fully in \kas. Table \ref{tab:sample} lists the archival
sample objects and the properties of the data. We note whether the
data are publicly available as of the time of submission of this paper
and list references where the data have appeared in the literature.

\begin{deluxetable}{rlcclrrcl}
\setlength{\tabcolsep}{0.1in}
\rotate
\tabletypesize{\small}
\tablewidth{0pt}
\tablecolumns{9}
\tablecaption{Archival data sample\label{tab:sample}}
\tablehead{
\colhead{MSX\,LMC\tablenotemark{a}} &
\colhead{SIMBAD Name\tablenotemark{b}} &
\colhead{RA (J2000)\tablenotemark{c}} &
\colhead{Dec (J2000)\tablenotemark{c}} &
\colhead{Target Name\tablenotemark{d}} &
\colhead{PID\tablenotemark{e}} &
\colhead{AORKEY\tablenotemark{f}} &
\colhead{Public?} &
\colhead{Ref.\tablenotemark{g}} \\
}
\startdata
21     & IRAS 05053-6659          &  05 05 20.35 & -66 55 06.6  & Cluster-013                      &  40650 &  23893504 &  N &          \\
43     & HV 888                   &  05 04 14.11 & -67 16 14.5  & HV 888                           &    200 &   6015488 &  Y & S08      \\
44     & IRAS 05112-6755          &  05 11 10.42 & -67 52 10.6  & TRM4                             &   3505 &  12939008 &  Y & Z06      \\
46     & LHA 120-N 17A            &  05 03 54.53 & -67 18 48.6  & SSTISAGE1C J050354.56-671848.5   &  40159 &  24317696 &  Y &          \\
47     & MXS LMC 47               &  05 11 13.85 & -67 36 16.2  & TRM24                            &   3505 &  12937984 &  Y & Z06      \\
80     & EQ 051005.7-685634       &  05 09 51.70 & -68 53 05.6  & lmc-hii-86                       &  40159 &  22470400 &  N &          \\
134    & HD 269006                &  05 02 07.39 & -71 20 13.1  & HD269006                         &   1404 &   9108736 &  Y &          \\
196    & IRAS 05125-7035          &  05 12 00.82 & -70 32 24.0  & LI-LMC 0603                      &   1094 &   6078464 &  Y &          \\
198    & HD 269211                &  05 12 30.17 & -70 24 22.3  & HD 269211                        &  30869 &  19151360 &  Y &          \\
215    & LHA 120-N 113A           &  05 13 21.74 & -69 22 39.4  & Cluster-005                      &  40650 &  23883008 &  N &          \\
219    & MSX LMC 219              &  05 11 19.46 & -68 42 27.7  & MSX051119.5-684227               &   3505 &  12939264 &  Y & Z06      \\
223    & LI-LMC 623               &  05 12 51.05 & -69 37 50.5  & MSX051250.8-693749               &   3505 &  12938752 &  Y & Z06      \\
262    & HD 34664                 &  05 13 52.99 & -67 26 54.6  & HD 34664                         &  30869 &  19149056 &  Y &          \\
283    & IRAS 05128-6455          &  05 13 04.56 & -64 51 40.3  & 05128-6455                       &    200 &   6024192 &  Y & S08      \\
307    & IRAS 05190-6748          &  05 18 56.28 & -67 45 04.7  & TRM20                            &   3505 &  12938496 &  Y & Z06      \\
318    & IRAS 05195-6911          &  05 19 12.31 & -69 09 06.5  & Cluster-002                      &  40650 &  23884288 &  N &          \\
323    & ARDB 184                 &  05 16 31.80 & -68 22 09.1  & LHA 120-S 93                     &  30869 &  19149312 &  Y &          \\
341    & MSX LMC 341              &  05 21 00.43 & -69 20 55.3  & MSX052100.5-692054               &   3505 &  12937728 &  Y & Z06      \\
349    & MSX LMC 349              &  05 17 26.98 & -68 54 58.7  & MSX051726.9-685458               &   3505 &  12938240 &  Y & Z06      \\
356    & 2MASS J05190229-6938033  &  05 19 02.28 & -69 38 03.5  & Cluster-017                      &  40650 &  23891456 &  N &          \\
398    & IRAS 05182-7117          &  05 17 34.61 & -71 14 57.5  & Cluster-016                      &  40650 &  23891968 &  N &          \\
441    & MSX LMC 441              &  05 24 38.71 & -70 23 56.8  & MSX052438.7-702357               &   3505 &  12936448 &  Y & Z06      \\
461    & LHA 120-N 132E           &  05 24 19.30 & -69 38 49.6  & IRAS 05247-6941\tablenotemark{h} &   1094 &   6076928 &  Y &          \\
464    & [HS66] 272               &  05 24 13.37 & -68 29 58.9  & HS 272                           &  40159 &  22425088 &  N &          \\
468    & BSDL 1469                &  05 22 53.28 & -69 51 10.4  & Cluster-024                      &  40650 &  23887872 &  N &          \\
500    & LI-LMC 861               &  05 21 29.69 & -67 51 07.2  & IRAS 05216-6753                  &   1094 &   6076160 &  Y &          \\
501    & NGC 1936                 &  05 22 12.55 & -67 58 31.8  & Cluster-006                      &  40650 &  23882496 &  N &          \\
560    & IRAS 05300-6651          &  05 30 03.86 & -66 49 24.2  & 05300-6651                       &    200 &   6024704 &  Y & S08      \\
581    & MSX LMC 581              &  05 26 46.63 & -68 48 46.8  & Cluster-002                      &  40650 &  23884288 &  N &          \\
596    & IRAS 05311-6836          &  05 30 54.36 & -68 34 27.8  & Cluster-006                      &  40650 &  23882496 &  N &          \\
601    & OGLE J052650.96-693136.8 &  05 26 50.83 & -69 31 36.8  & MSX052650.9-693136               &   3505 &  12935424 &  Y & Z06      \\
635    & IRAS 05278-6942          &  05 27 24.12 & -69 39 45.0  & MSX052724.3-693944               &   3505 &  12932864 &  Y & Z06      \\
640    & LHA 120-N 129            &  05 22 24.94 & -69 42 32.4  & Cluster-010                      &  40650 &  23895040 &  N &          \\
646    & [MLD95] LMC 1-289        &  05 30 47.88 & -71 07 55.2  & LMC 1-289                        &  30869 &  19150336 &  Y &          \\
651    & IRAS 05310-7110          &  05 30 20.16 & -71 07 48.4  & Cluster-042                      &  40650 &  23894784 &  N &          \\
690    & MSX LMC 690              &  05 32 14.54 & -71 13 28.2  & Cluster-026                      &  40650 &  23886848 &  N &          \\
692    & MSX LMC 692              &  05 28 46.63 & -71 19 12.7  & IRAS 05295-7121                  &   3505 &  12929024 &  Y & Z06      \\
733    & IRAS 05348-7024          &  05 34 15.98 & -70 22 52.7  & 05348-7024                       &    200 &   6024448 &  Y & S08      \\
766    & LHA 120-N 150            &  05 33 42.22 & -68 46 00.8  & Cluster-014                      &  40650 &  23892992 &  N &          \\
771    & MSX LMC 771              &  05 32 38.59 & -68 25 22.1  & MSX053238.7-682522               &   3505 &  12931328 &  Y &          \\
805    & LI-LMC 1163              &  05 32 35.62 & -67 55 08.8  & HV 996                           &    200 &   6015744 &  Y & S08      \\
811    & MSX LMC 811              &  05 32 51.34 & -67 06 51.8  & 05329-6708                       &    200 &   6023168 &  Y & S08      \\
886    & IRAS 05389-6922          &  05 38 33.96 & -69 20 31.6  & IRAS 05389-6922                  &   1094 &   6076416 &  Y &          \\
887    & IRAS 05406-6924          &  05 40 13.33 & -69 22 46.5  & HD 38489                         &  30869 &  19149824 &  Y &          \\
936    & IRAS 05402-6956          &  05 39 44.86 & -69 55 18.1  & 05402-6956                       &    200 &   6020608 &  Y & S08      \\
1117   & IRAS 04498-6842          &  04 49 41.47 & -68 37 51.2  & IRAS 04498-6842                  &   1094 &   6076672 &  Y &          \\
1130   & IRAS 04496-6958          &  04 49 18.50 & -69 53 14.3  & IRAS 04496-6958                  &   1094 &   9069312 &  Y & S06      \\
1171   & IRAS 04545-7000          &  04 54 10.08 & -69 55 58.4  & 04545-7000                       &    200 &   6020352 &  Y & S08      \\
1183   & BSDL 126                 &  04 51 53.66 & -69 23 28.3  & Cluster-003                      &  40650 &  23883776 &  N &          \\
1184   & IRAS 04530-6916          &  04 52 45.67 & -69 11 49.6  & 04530-6916                       &    200 &   6023936 &  Y & S08      \\
1190   & IRAS 04516-6902          &  04 51 29.02 & -68 57 49.7  & 04516-6902                       &    200 &   6020096 &  Y & S08      \\
1191   & WOH S 60                 &  04 53 30.86 & -69 17 49.9  & GV 60                            &  40159 &  22402560 &  Y &          \\
1192   & IRAS 04509-6922          &  04 50 40.46 & -69 17 31.9  & 04509-6922                       &    200 &   6022400 &  Y & S08      \\
1193   & MSX LMC 1193             &  04 50 23.40 & -69 37 56.6  & IRAS04506-6942                   &  30788 &  19009792 &  Y &          \\
1207   & LHA 12-N 89              &  04 55 06.58 & -69 17 08.5  & Cluster-022                      &  40650 &  23888896 &  N &          \\
1225   & MSX LMC 1225             &  04 57 47.98 & -66 28 44.8  & Cluster-007                      &  40650 &  23881984 &  N &          \\
1247   & PGMW 3123                &  04 56 47.04 & -66 24 31.3  & lmc-hii-34                       &  40159 &  22469632 &  Y &          \\
1278   & IRAS 05009-6616          &  05 01 04.42 & -66 12 40.3  & IRAS 05009-6616                  &   3505 &  12929536 &  Y & Z06      \\
1296   & HD 32364                 &  04 57 14.33 & -68 26 30.5  & HD 32364                         &  30869 &  19150592 &  Y &          \\
1302   & IRAS 04589-6825          &  04 58 46.27 & -68 20 42.7  & IRAS04589-6825                   &  30788 &  19010304 &  Y &          \\
1438   & HD 269997                &  05 41 21.19 & -69 04 38.6  & HD 269997                        &  40159 &  22440704 &  N &          \\
1453   & IRAS 05506-7053          &  05 49 56.54 & -70 53 11.8  & IRAS05506-7053                   &  30788 &  19005952 &  Y &          \\
1651   & MSX LMC 1651             &  06 02 45.10 & -67 22 43.3  & LI-LMC 1817\tablenotemark{h}     &   1094 &   6078208 &  Y &          \\
\enddata
\tablenotetext{a}{MSX\,LMC identifier \eganp.}
\tablenotetext{b}{Object name in SIMBAD (simbad.u-strasbg.fr/sim-fid.pl).}
\tablenotetext{c}{MSX\,LMC source position from SIMBAD}
\tablenotetext{d}{Object name as given in the headers of the data.}
\tablenotetext{e}{Spitzer program identifier.}
\tablenotetext{f}{Unique identifier for the astronomical observation request (AOR).}
\tablenotetext{g}{References. -- S08: \protect\citealt{slo08}; S06: \protect\citealt{spe06}; Z06: \protect\citealt{zij06}.}
\tablenotetext{h}{For these objects, the target name does not match the position observed, possibly due to misidentification or peakup failure. In all cases, the position observed matches the position of the MSX\,LMC source.}
\end{deluxetable}

All of the targets except MSX\,LMC~500 were observed with the
low-resolution (SL and LL) modules of the IRS. They extend from 5.2 --
14~\micron\ (SL) and 14 -- 38~\micron\ (LL), with resolving powers of
64 -- 128.  The SL module has a slit width of 3.6 -- 3.7\arcsec, and
the LL module a slit width of 10.5 -- 10.7\arcsec. The raw data were
processed through the \spitzer\ pipeline versions S15.3 or later.
After pipeline processing, the basic calibrated data (BCDs) were
cleaned for rogue pixels, using the IRSCLEAN
software\footnote{IRSCLEAN was written by the IRS GTO team (G.\ Sloan,
  D.\ Devost, \& B.\ Sargent). It is distributed by the
  \spitzer\ Science Center at Caltech.}.  For objects observed in
staring mode (all except MSX\,LMC~1247), spectra were extracted using
the SMART software\footnote{SMART was developed by the IRS Team at
  Cornell University and is available through the \spitzer\ Science
  Center at Caltech.} \citep{hig04}.  Where multiple exposures were
obtained, the two-dimensional spectral images (BCDs) of the multiple
exposures were median-combined.  Off-source regions of each spectral
image served as sky background for the spectral image obtained with
each module.  Spectra were extracted and the fluxes were calibrated in
SMART using the default point source extraction apertures.  The
modules at each nod position were merged and the edges and overlapping
regions of the modules were trimmed.  The spectra from the two nod
positions were then averaged to produce the final
spectrum. Uncertainty images are provided by the SSC pipeline and
propagated through SMART to produce the uncertainties on the final
spectra.

For 11 sources, a flux jump of $\gtrsim$5\% was observed between the
SL and LL module spectra ($\sim$14~\micron). These were due to
extended emission falling in the IRS slits and the differing widths of
the SL and LL slits (see, e.g., \buc). In most cases, the LL spectrum
was scaled down to match the SL spectrum, but for the sources
identified as \htwo\ regions the SL was scaled up to match the LL
spectrum (\S\ref{sec:res}).

One source, MSX\,LMC~1247, was observed in mapping mode. For each
module and order (SL1, SL2, LL1, and LL2), a spectral cube was
constructed using the Cubism software \citep{smi07}. Individual BCDs
were background subtracted, using the corresponding off-source module,
and corrected for bad pixels before the cubes were built.
One-dimensional spectra were then extracted by integrating over a
10\arcsec\ region around the source in each cube.

MSX\,LMC~500 was observed with the SL module and both high resolution
modules (SH and LH) of the IRS. The high resolution modules produce
spectra covering 10 -- 19~\micron\ (SH) and 19 -- 37~\micron\ (LH),
with a resolving power of $\sim$600. The reduction of the high
resolution modules was carried out in the same manner as for the low
resolution data, except for the background subtraction and the
extraction aperture. No separate sky observations were taken for these
data, so the sky background as a function of wavelength was estimated
using SPOT and subtracted from the spectra after extraction.  The high
resolution spectra were extracted and fluxes calibrated using the
default full extraction aperture in SMART.  There are mismatches in
flux between the orders in the LH spectrum which are most likely due
to poor sky subtraction. These were rectified by scaling the longer
wavelength order to match the shorter wavelength order.  No such
mismatches are observed in the SH spectrum, though the SH spectrum was
divided by a factor of 1.2 to match the SL spectrum.

\section{SPECTRA AND CLASSIFICATIONS} \label{sec:res}

In this section we present the previously unpublished spectra and
classify them according to their spectral features.  Table
\ref{tab:class} lists the predicted JHK8 object classes from \kas, the
dominant spectral features and luminosity we determined from the IRS
spectrum, and the `actual' (most probable) object class based on the
available information, for the 44 archival sample objects for which
data are published or available (see \S\ref{sec:obs} and Table
\ref{tab:sample}). We note that, for the previously published spectra,
our classifications agree with those of the publishing authors in all
cases except two (MSX\,LMC~1190 and 1192; see \S\ref{subsubsec:res_oagb}),
for which we adopt the classification of \citet{slo08}.  Infrared (1
-- 100~\micron) luminosities were estimated from the IRS spectra and
2MASS photometry using the method outlined in \buc. Typical
uncertainties in the luminosity for the stellar sources are up to
15\%, ignoring the effects of variability, but may be a factor of a
few for the compact \htwo\ regions, where the bulk of the luminosity
comes from wavelengths longer than the IRS spectral range.  In the
following subsections, we discuss the spectral classifications
arranged according to object class.

\begin{deluxetable}{clllcll}
\setlength{\tabcolsep}{0.15in}
\tabletypesize{\small}
\tablewidth{0pt}
\tablecolumns{7}
\tablecaption{Summary of classifications \label{tab:class}}
\tablehead{
\colhead{MSX\,LMC} &
\colhead{JHK8\tablenotemark{a}} &
\colhead{Spectral\tablenotemark{b}} &
\colhead{Ref.\tablenotemark{c}} &
\colhead{Fig.} &
\colhead{\protect\lir} &
\colhead{Object\tablenotemark{b}} \\
\colhead{number} &
\colhead{Class} &
\colhead{Features} &
\colhead{} &
 &
\colhead{($\times 10^{4}$~\protect\lsun)} &
\colhead{Class}
}
\startdata
\phm{1}283   &  -              & O-RICH                  &  S08         &  \nodata                 & \nodata             & O AGB          \\ 
      1190   &  -              & O-RICH                  &  S08         &  \nodata                 & \nodata             & O AGB          \\ 
      1192   &  GMV:           & O-RICH                  &  S08         &  \nodata                 & \nodata             & O AGB          \\ 
\\
\phm{1}811   &  C AGB:         & O-RICH\tablenotemark{d} &  S08         &  \nodata                 & \nodata             & OH/IR          \\ 
\phm{1}936   &  C AGB          & O-RICH\tablenotemark{d} &  S08         &  \nodata                 & \nodata             & OH/IR          \\ 
      1171   &  C AGB:         & O-RICH\tablenotemark{d} &  S08         &  \nodata                 & \nodata             & OH/IR          \\ 
\\
\phm{12}43   &  RSG            & O-RICH                  &  S08         &  \nodata                 & \nodata             & RSG            \\ 
\phm{1}461   &  RSG/GMV        & O-RICH\tablenotemark{e} &  this paper  &  \protect\ref{fig:specO} & 18:                 & RSG            \\
\phm{1}805   &  RSG            & O-RICH                  &  S08         &  \nodata                 & \nodata             & RSG            \\ 
\phm{1}886   &   -             & O-RICH                  &  this paper  &  \protect\ref{fig:specO} & 10                  & RSG            \\ 
      1117   &  RSG/GMV        & O-RICH                  &  this paper  &  \protect\ref{fig:specO} & 11                  & RSG            \\ 
      1191   &  RSG            & O-RICH\tablenotemark{f} &  this paper  &  \protect\ref{fig:specO} & 7.0                 & RSG            \\ 
\phm{1}500   &  \HII           & O-RICH\tablenotemark{f} &  this paper  &  \protect\ref{fig:specO} & 16                  & RSG?           \\ 
 \\
\phm{1}262   &  C/O AGB:       & O-RICH                  &  this paper  &  \protect\ref{fig:specO} & 8.3                 & B[e]           \\ 
\phm{1}323   &  -              & O-RICH                  &  this paper  &  \protect\ref{fig:specO} & 1.7                 & B[e]           \\
\phm{1}887   &  C/O AGB:       & O-RICH                  &  this paper  &  \protect\ref{fig:specO} & 2.6                 & B[e]           \\ 
\phm{1}134   &  -              & PAH/O-RICH              &  this paper  &  \protect\ref{fig:specO} & 11                  & B[e]?          \\ 
 \\
\phm{12}44   &  C AGB          & C-RICH                  &  Z06         &  \nodata                 & \nodata             & C AGB          \\ 
\phm{12}47   &  C AGB:         & C-RICH                  &  Z06         &  \nodata                 & \nodata             & C AGB          \\ 
\phm{1}196   &  -              & C-RICH                  &  this paper  &  \protect\ref{fig:specC} & 2.3                 & C AGB          \\ 
\phm{1}219   &  \HII:          & C-RICH                  &  Z06         &  \nodata                 & \nodata             & C AGB          \\ 
\phm{1}223   &  -              & C-RICH                  &  Z06         &  \nodata                 & \nodata             & C AGB          \\ 
\phm{1}307   &  C AGB:         & C-RICH                  &  Z06         &  \nodata                 & \nodata             & C AGB          \\ 
\phm{1}341   &  C AGB:         & C-RICH                  &  Z06         &  \nodata                 & \nodata             & C AGB          \\ 
\phm{1}349   &  -              & C-RICH                  &  Z06         &  \nodata                 & \nodata             & C AGB          \\ 
\phm{1}441   &  -              & C-RICH                  &  Z06         &  \nodata                 & \nodata             & C AGB          \\ 
\phm{1}560   &  C AGB          & C-RICH                  &  S08         &  \nodata                 & \nodata             & C AGB          \\ 
\phm{1}601   &  -              & C-RICH                  &  Z06         &  \nodata                 & \nodata             & C AGB          \\ 
\phm{1}635   &  -              & C-RICH                  &  Z06         &  \nodata                 & \nodata             & C AGB          \\ 
\phm{1}692   &  C AGB          & C-RICH                  &  Z06         &  \nodata                 & \nodata             & C AGB          \\ 
\phm{1}733   &  C AGB:         & C-RICH                  &  S08         &  \nodata                 & \nodata             & C AGB          \\ 
      1130   &  C/O AGB        & C-RICH                  &  S06         &  \nodata                 & \nodata             & C AGB          \\ 
      1278   &  C AGB:         & C-RICH                  &  Z06         &  \nodata                 & \nodata             & C AGB          \\ 
      1302   &  \HII?          & C-RICH?                 &  this paper  &  \protect\ref{fig:specC} & 0.2:                & C AGB?         \\ 
      1453   &  C AGB:         & C-RICH                  &  this paper  &  \protect\ref{fig:specC} & 1.7                 & C AGB          \\ 
      1651   &  C AGB:         & C-RICH                  &  this paper  &  \protect\ref{fig:specC} & 1.2                 & C AGB          \\ 
 \\
\phm{1}198   &  {\protect\HII} & PAH                     &  this paper  &  \protect\ref{fig:specH} & 2.6                & {\protect\HII}  \\ 
\phm{1}646   &  {\protect\HII}:& PAH                     &  this paper  &  \protect\ref{fig:specH} & 1.7                & {\protect\HII}  \\ 
      1247   &  {\protect\HII} & PAH                     &  this paper  &  \protect\ref{fig:specH} & 10                 & {\protect\HII}  \\ 
      1296   &  {\protect\HII} & PAH                     &  this paper  &  \protect\ref{fig:specH} & 5.4                & {\protect\HII}  \\ 
 \\
\phm{12}46   &  -              & PAH                     &  this paper  &  \protect\ref{fig:specY} & 1.6                & YSO?            \\ 
\phm{1}771   &  C AGB          & PAH?                    &  this paper  &  \protect\ref{fig:specY} & 2.5                & YSO?            \\ 
      1184   &  C AGB          & PAH                     &  S08         &  \nodata                 & \nodata            & YSO             \\ 
 \\
1193\tablenotemark{g} & C AGB  & \nodata                 &  \nodata     &  \nodata                 & \nodata            &                 \\ 
\enddata
\tablenotetext{a}{\protect\kas\ color-color classification.}
\tablenotetext{b}{Spectral features refers to the dominant dust
  features in the IRS spectrum which are usually unambiguous. The
  object class refers to our classification of the object, based on
  its spectral features and other information, which is somewhat more
  uncertain.}
\tablenotetext{c}{S08: \protect\citealt{slo08}; S06: \protect\citealt{spe06}; Z06: \protect\citealt{zij06}.}
\tablenotetext{d}{These objects show silicate self-absorption.}
\tablenotetext{e}{The LL spectrum is contaminated by emission from a nearby source blended with that of MSX\,LMC~461. See text for details.}
\tablenotetext{f}{Surrounded by \protect\htwo\ region.}
\tablenotetext{g}{This object was not detected due to mispointing.}
\end{deluxetable}

\subsection{Oxygen-rich objects} \label{subsec:res_orich}

\begin{figure}
\epsscale{0.6}
\plotone{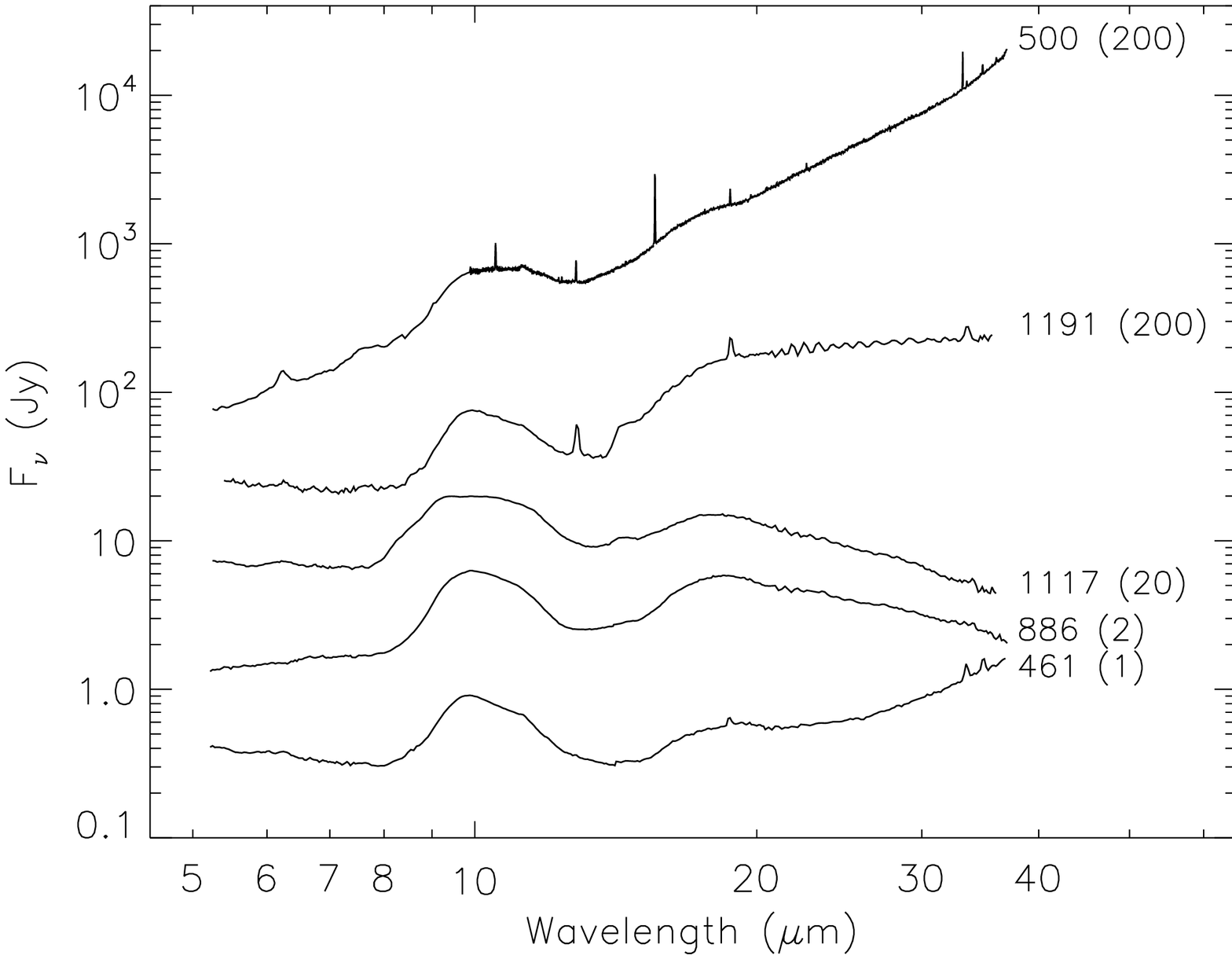}

\plotone{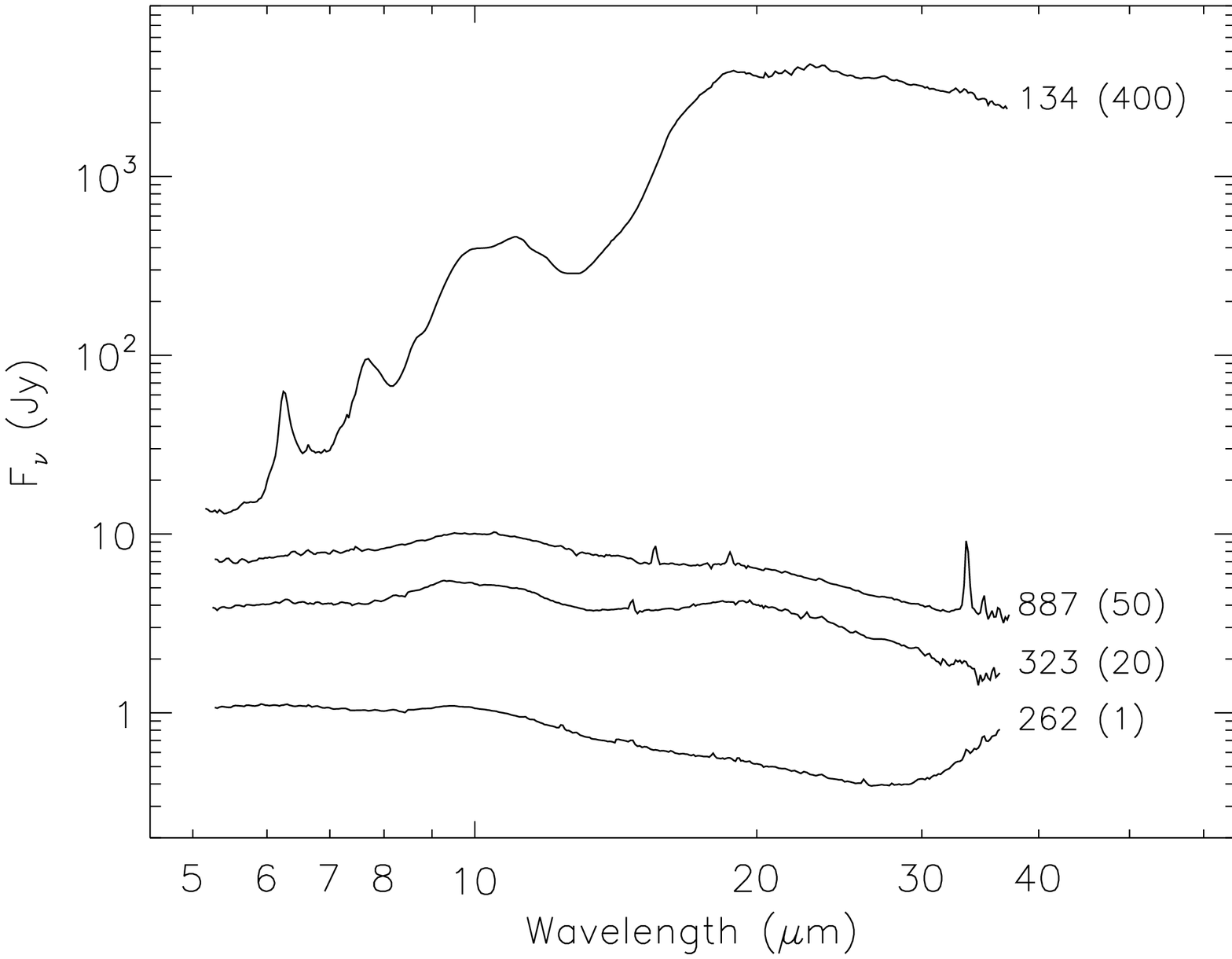}
\caption{IRS spectra of objects spectroscopically identified as
  oxygen-rich. \emph{Top panel:} Red supergiants. \emph{Bottom panel:}
  Known and candidate B[e] stars. The MSX\,LMC number of each source
  is listed by the spectrum and, for clarity, spectra are multiplied
  by the factor indicated in parentheses. Where a flux disparity
  occurs between the SL and LL modules (see text) the LL module has
  been scaled to the SL module flux.  \label{fig:specO}}
\epsscale{1.0}
\end{figure}

We group the classes of object that have oxygen-rich dust chemistry
together, and discuss them in the following subsubsections.  Seventeen
objects were identified as having oxygen-rich dust chemistry from
their IRS spectra.  Figure \ref{fig:specO} shows the 9 spectra that
have not previously been published.  Stars with oxygen-rich
circumstellar dust show broad silicate features at 9.7 and 18~\micron,
usually in emission.  While oxygen-rich circumstellar dust is easy to
identify on the basis of its mid-IR spectral features, classifying the
nature of a visually obscured star with an O-rich circumstellar
envelope as an AGB star or a red supergiant requires an assessment of
its bolometric luminosity.  The O-rich objects in this sample include
3 O-rich AGB stars, 3 OH/IR stars, 7 red supergiants (RSGs), one of
which is a tentative classification, and 4 dusty, early-type stars. We
now discuss each of these subcategories describing individual objects
and spectra where appropriate.

\subsubsection{O-rich asymptotic giant branch stars} \label{subsubsec:res_oagb}

Three of the O-rich stars listed in Table 2, MSX\,LMC~283, 1190, and
1192, were classified as AGB stars by \citet{slo08}.  MSX\,LMC~283 has
an IR luminosity 1.4$ \times 10^{4}$~\lsun, and is clearly an AGB
star. MSX\,LMC~1190 and 1192 both have IR luminosities $\sim7 \times
10^{4}$~\lsun, just above the classical limit for AGB stars,
suggesting they are RSGs; however, we adopt the classification of
\citet{slo08}, who class these two stars as O-rich AGB stars on the
basis of their amplitudes of K-band variability.

\subsubsection{OH/IR stars}

Three oxygen-rich sources show silicate self-absorption at
9.7~\micron. All three, MSX\,LMC~811, 936, and 1171, are classified as
OH/IR stars --- i.e., O-rich AGB stars with especially optically thick
circumstellar envelopes and that display OH maser emission --- by
\citet{slo08}.

\subsubsection{Red supergiants} \label{subsubsec:res_rsg}

Seven objects were classified as red supergiants (RSGs) on the basis of
their \spitzer\ spectral features and luminosity. These objects all
have estimated luminosities $\gtrsim$7.0$\times
10^{4}$~\lsun. \\ 
\emph{MSX\,LMC~43} and \emph{MSX\,LMC~805:} These
stars are classified as RSGs by \citet{slo08}.\\ 
\emph{MSX\,LMC~461:} The rising SED of RSG MSX\,LMC~461 above
$\sim$20~\micron\ is likely due to a contaminating source in the LL
module slits.  The 24~\micron\ SAGE image shows that MSX\,LMC~461 is
in a cluster at the center of an \htwo\ region, with 3 bright
24~\micron\ sources within $\sim$25\arcsec. The nearest of these is
separated by 11\arcsec\ according to the SAGE 24~\micron\ catalog. The
24~\micron\ flux of the contaminating source is 347~mJy, making it
slightly brighter than MSX\,LMC~461 (335~mJy), but it does not appear
in the IRAC catalog or images, indicating it is very red.  An overlay
of the observed slit positions on the 24~\micron\ image indicates this
source fell in the LL slits.  Examination of the LL1 and LL2 spectral
images reveals that two blended sources separated by $\sim$2 pixels
(5.1\arcsec/pixel) can be distinguished at the shorter wavelengths
($\lesssim$27~\micron).  As these sources are heavily blended, no
attempt was made to separate their spectra.  In addition, the SAGE
70~\micron\ image shows diffuse emission due to cold dust in the
surrounding \htwo\ region, which may also have contaminated the longer
wavelengths of the IRS spectrum.  Therefore, the luminosity for
MSX\,LMC~461 was estimated using only the SL module spectra (5.2 --
14~\micron). While the luminosity is therefore somewhat uncertain, it
is still sufficiently high ($1.8 \times 10^5$:~\lsun) to confidently
classify it as an RSG. The LL spectrum was scaled by 0.94 to match the
SL spectrum. \\
\emph{MSX\,LMC~886:} Although the JHK8 colors and magnitudes of
MSX\,LMC~886 are more similar to O-rich AGB stars than RSGs, we
estimate its IR luminosity to be 10$^{5}$~\lsun, and \citet{woo92}
identify it as a probable core He-burning supergiant as it shows no
large-amplitude variability.\\
\emph{MSX\,LMC~1117:} This star has a luminosity of $1.1 \times
10^{5}$~\lsun, placing it clearly in the RSG category. \\
\emph{MSX\,LMC~1191:} We classify this star as a RSG due to its
luminosity of $7 \times 10^{4}$~\lsun.  This unusual spectrum, with a
flat mid-IR slope, is reminiscent of sources with dust disks (see
\citealt{kas06}) and, since these are of special interest,
MSX\,LMC~1191 warrants further investigation.  No contaminating source
is readily apparent in the spectral image of MSX\,LMC~1191. This
object appears point-like in the SAGE 24~\micron\ image, although
there is faint \neii, \siiia, and \siiib\ emission off-source and a
small SL/LL jump suggests extended emission is associated with the
source. The LL spectrum was scaled by 0.90 to match the SL
spectrum. \\
\emph{MSX\,LMC~500:} This star appears to be a mass-losing O-rich
evolved star within an \htwo\ region.  The spectrum clearly shows
silicate features in emission, but also has a rising continuum, and
features typical of \htwo\ regions, including PAH features at
6.2~\micron, 7.7~\micron, and 11.2~\micron, and emission lines \siv,
\neii, \neiiia, \siiia, \siiib, \silii, and \neiiib.  Although it is
not included in the SAGE 24 micron point source catalog, SAGE IRAC and
MIPS images show a bright point source surrounded by strong nebulosity
at the position of MSX\,LMC~500.  The contamination of this spectrum
by the \htwo\ region prevents certain identification of this object as
an O-rich evolved star as opposed to a massive YSO, but here we
tentatively classify it as a RSG (rather than an O-rich AGB star), on
the basis of its high luminosity ($1.6 \times 10^{5}$~\lsun). We note
that this classification is uncertain, as the luminosity includes a
significant contribution from the contaminating \htwo\ region. 

\subsubsection{Dusty, early-type stars}

Three of the objects showing O-rich dust are known B[e] supergiants
(\kas). These three, MSX LMC 262, 323, and 887, all show flat IRS
spectra with (relatively weak) silicate
features in emission (Figure \ref{fig:specO}). In addition, the source
MSX\,LMC~134 -- which displays perhaps the most peculiar spectrum
among the IR sources considered here -- is associated with a luminous
O/B star. These four sources will be discussed in more detail in a
forthcoming paper on \spitzer\ observations of dusty early-type stars
in the Magellanic Clouds (J. Kastner et al.\ 2009, in preparation). \\
\emph{MSX\,LMC~262:} The spectrum of this star (HD\,34664) has slowly
falling (blue) continuum and what appears to be weak silicate
features. This source is affected by contamination from diffuse line
emission, as is apparent in the spectral images, although this appears to
be effectively removed by the sky subtraction.  The rising continuum
above $\sim$29~\micron\ is attributable to a nearby contaminating
source, observed in IRAC and MIPS images, which is very red and
blended with MSX\,LMC~262. The LL spectrum was scaled by 0.96 to match the SL
spectrum.  \\
\emph{MSX\,LMC~323:} MSX\,LMC~323 (LHA\,120-S\,93) shows the strongest
silicate emission features among the 3 B[e] supergiants with spectra
in Figure 1. There is a small jump in flux density between the SL and
LL modules, consistent with faint diffuse emission in the LL module
spectral images; however no narrow emission lines are observed in the
sky or the stellar spectrum. The LL spectrum was scaled by 0.85 to
match the SL spectrum. \\
\emph{MSX\,LMC~887:} In addition to the broad silicate features, this
star (HD\,38489) shows narrow emission lines that we identify as
\neiiia\ and \siiia.  Examination of the spectral images indicates
that these lines arise from extended, diffuse emission and not from
the star itself. Also present in the images are emission lines of
\siv, \neii, \siiib, and \silii, which are too weak to show up in the
extracted stellar spectrum.  The spectrum of MSX\,LMC~887 also shows a
large jump in flux density (a factor of $\sim$5) between the SL and LL
modules, most likely due to extended emission apparent in the LL
module spectral images. The LL spectrum was scaled by 0.21 to match
the SL spectrum.  SAGE 8~\micron\ and 24~\micron\ images show this
star to be a point source surrounded by nebulous emission associated
with a \htwo\ region, consistent with the presence of the ionized gas
seen in the spectral images. \\
\emph{MSX\,LMC~134:} This star (HD\,269006) shows a peculiar spectrum
with very strong silicate emission feature, strong PAH emission at
6.2~\micron\ and 7.7~\micron, and a steeply rising continuum that
flattens above $\sim$20~\micron. This flattened spectral shape is
suggestive of a dust disk with a large central hole or gap. \\

\subsection{Carbon-rich AGB stars} \label{subsec:res_cagb}

\begin{figure}
\epsscale{0.6}
\plotone{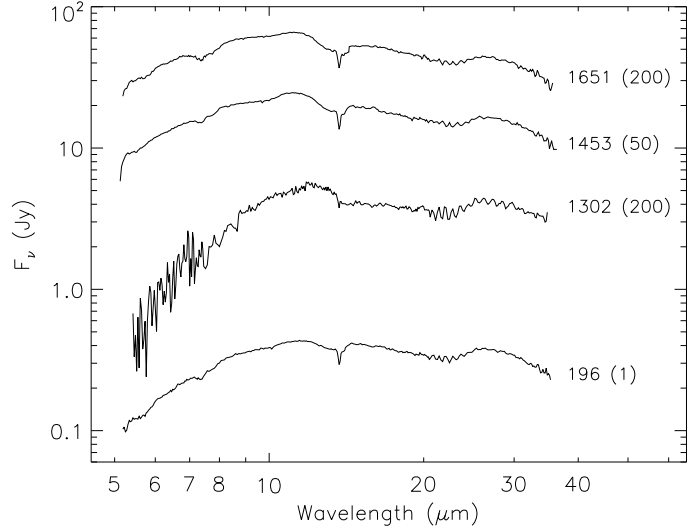}
\caption{IRS spectra of objects identified as C-rich AGB stars. The
  MSX\,LMC number of each source is listed by the spectrum and, for
  clarity, spectra are multiplied by the factor indicated in
  parentheses. Where a flux disparity occurs between the SL and LL
  modules (see text) the LL module has been scaled to the SL module
  flux. \label{fig:specC}} \epsscale{1.0}
\end{figure}

Nineteen of the objects in the archival sample are classified as
carbon-rich on the basis of their \spitzer\ IRS spectra.  They are
characterized by an absorption feature at 8~\micron\ due to
\ctwohtwo\ and \hcn, a SiC dust emission feature at 11.3~\micron, a
narrow absorption feature at 13.7~\micron\ due to \ctwohtwo\ gas, and
a broad emission feature around 26 -- 30~\micron\ attributed to
\mgs\ \citep{goe85,hon02}.  Figure \ref{fig:specC} shows the IRS
spectra of the 4 objects for which IRS data have not previously been
published (see Tables \ref{tab:sample} and \ref{tab:class}).  The
references for the remaining 15 are listed in Table
\ref{tab:class}. The LL spectrum of MSX\,LMC~1651 was scaled by 0.91
to match the SL spectrum.

The classification of MSX\,LMC~1302 is somewhat uncertain, due to the
relatively low signal-to-noise of the spectrum and the presence of a
large (factor $\sim$5) jump between the SL and LL modules. The LL
spectrum was scaled by a factor 0.19 to match the SL spectrum.  This
jump is most likely due to diffuse extended emission that is due to cool dust, as seen in the
SAGE 8.0~\micron\ and 24~\micron\ images as well as in the LL spectral
images. The LL images also reveal \siiib\ and
\silii\ emission surrounding the source, but this emission is too weak
to be apparent in the spectrum.

\subsection{Compact \protect\htwo\ regions} \label{subsec:res_hii}

\begin{figure}
\epsscale{0.6}
\plotone{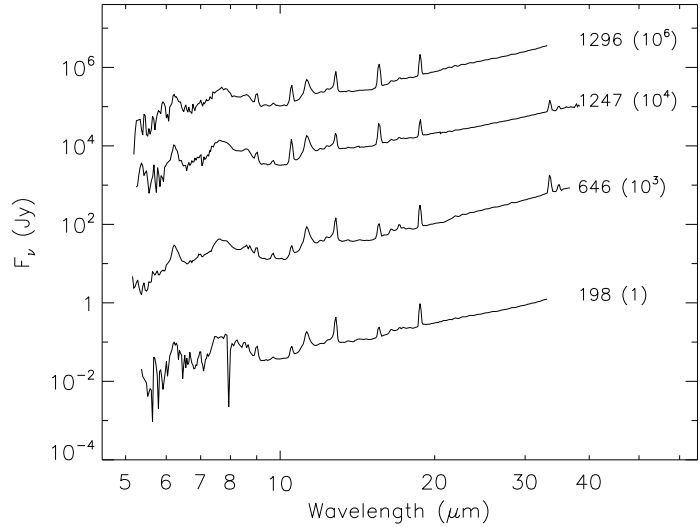}
\caption{IRS spectra of objects identified as
  \protect\htwo\ regions. The MSX\,LMC number of each source is listed
  by the spectrum and, for clarity, spectra are multiplied by the
  factor indicated in parentheses. Where a flux jump occurs between
  the SL and LL modules, the SL spectrum was scaled to match the LL
  spectrum (see text). \label{fig:specH}} \epsscale{1.0}
\end{figure}

Figure \ref{fig:specH} shows the spectra of the 4 objects
identified as compact \htwo\ regions or candidate compact
\htwo\ regions on the basis of their IRS spectra.  These sources are
characterized by very red continua, strong PAH emission features at
6.2, 7.7, 8.6, 11.3, and 12.7~\micron, silicate absorption at
9.7~\micron, and the following lines of ionized species: \ariii, \siv,
\neii, \neiiia, \siiia, \siiib, \silii, and \neiiib.  All these
sources except MSX\,LMC~1247 
also show a flux difference at 14~\micron\ between the SL and LL
modules, due to their extended nature; in Figure \ref{fig:specH}, the
SL spectra have been scaled by factors 1.4 -- 2.6 to match the LL
fluxes. No flux jump is seen in MSX\,LMC~1247 as it was observed in
mapping mode.

\subsection{Young Stellar Objects and Candidates} \label{subsec:res_yso}

\begin{figure}
\epsscale{0.6}
\plotone{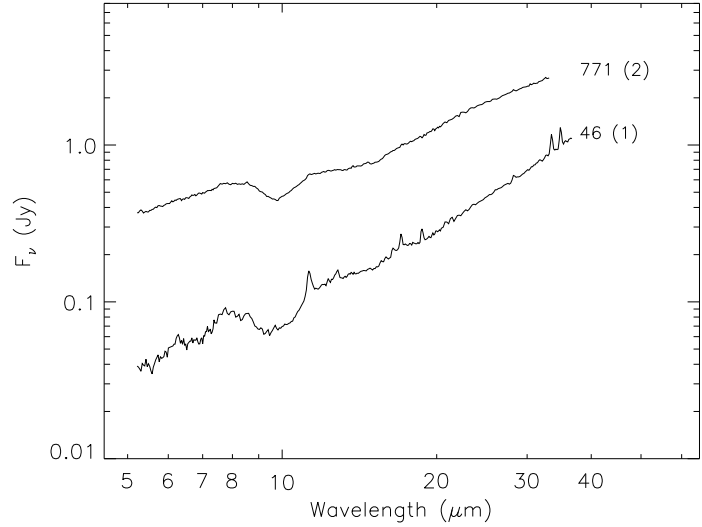}
\caption{IRS spectra of candidate young stellar objects. The MSX\,LMC
  number of each source is listed by the spectrum and, for clarity,
  spectra are multiplied by the factor indicated in parentheses. The
  LL spectra were scaled to match the SL spectra (see
  text). \label{fig:specY}} \epsscale{1.0}
\end{figure}

\begin{figure}
\epsscale{0.6}
\plotone{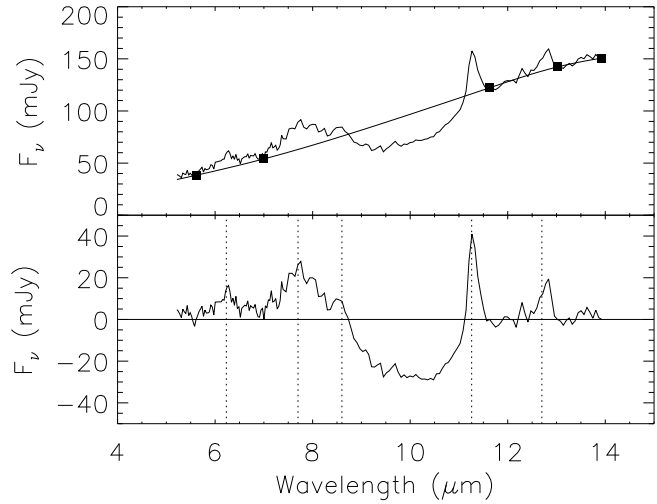}
\caption{SL module spectrum of MSX\,LMC~46. {\it Top panel:} the
  spline {\it (thin line)} fitted to the continuum points {\it
    (squares)}. {\it Bottom:} The spline-subtracted spectrum,
  highlighting the PAH emission features at wavelengths 6.23, 7.70,
  8.60, 11.26, and 12.70~\micron\ {\it (dotted
    lines)}. \label{fig:yso46}} \epsscale{1.0}
\end{figure}

One star in the sample, MSX\,LMC~1184, has been classified as a young
stellar object (YSO) based on its optical and IR properties
\citep{slo08}.  It shows a red continuum, similar to \htwo\ regions,
with PAH emission and silicate absorption features, but lacks the
narrow ionized line emission observed in \htwo\ regions.  We find two
more objects in the sample with somewhat similar IR spectra, and group
them here with 1184, though we note their classification as candidate
YSOs is very tentative. \\
\emph{MSX\,LMC~46:} The spectrum of this star resembles the spectrum
of YSO MSX\,LMC~1184 presented by \citet{slo08}, and shows much weaker
line emission and PAH features than the \htwo\ regions, which have
similar red continua (Figure \ref{fig:specY}). The LL spectrum was
scaled by 0.74 to match the SL spectrum. Figure \ref{fig:yso46} shows
the SL spectrum of MSX\,LMC~46, with a spline-fitted continuum
subtracted to highlight the PAH features and silicate absorption.  It
is possible this object is a high-mass protostar in transition from
the embedded young stellar object (YSO) stage to the zero-age main
sequence. \\
\emph{MSX\,LMC~771:} MSX\,LMC~771 shows a similar continuum shape to
MSX\,LMC~1184 (Figure \ref{fig:specY}), but the nature of this source
is uncertain. The only clear spectral feature in this spectrum is
silicate absorption at 9.7~\micron. The silicate absorption profile is
more sharply peaked, and the PAH features are significantly weaker
than in MSX\,LMC~1184 and 46.  This source may be a compact
\htwo\ region, though it lacks the lines of ionized species normally
observed in the spectra of \htwo\ regions.  Weak emission lines
\neiiia, \siiia, \siiib, and \silii, and lines at 16.4~\micron\ and
17.1~\micron, are apparent in the spectral images, but do not appear
in the extracted spectrum. SAGE 8 and 24~\micron\ images show this
object to be a bright point source located within a nebulous region.
It is possible this source is an embedded YSO. The LL spectrum was
scaled by 0.94 to match the SL spectrum. \\
\emph{MSX\,LMC~1184:} This star has been identified as a young stellar
object (YSO; \citealt{slo08,van05}), based on its optical and IR
spectral features and its IR colors.  This source shows a red
continuum, with PAH emission features and silicate absorption at
9.7~\micron.

\subsection{Unclassified sources} \label{subsec:res_unc}

The observation of MSX\,LMC~1193 was mispointed; therefore this object
was not detected and could not be classified.

\section{CHECKING AND IMPROVING THE JHK8 CLASSIFICATIONS} \label{sec:dis}

\begin{figure} 
\epsscale{0.7}
\plotone{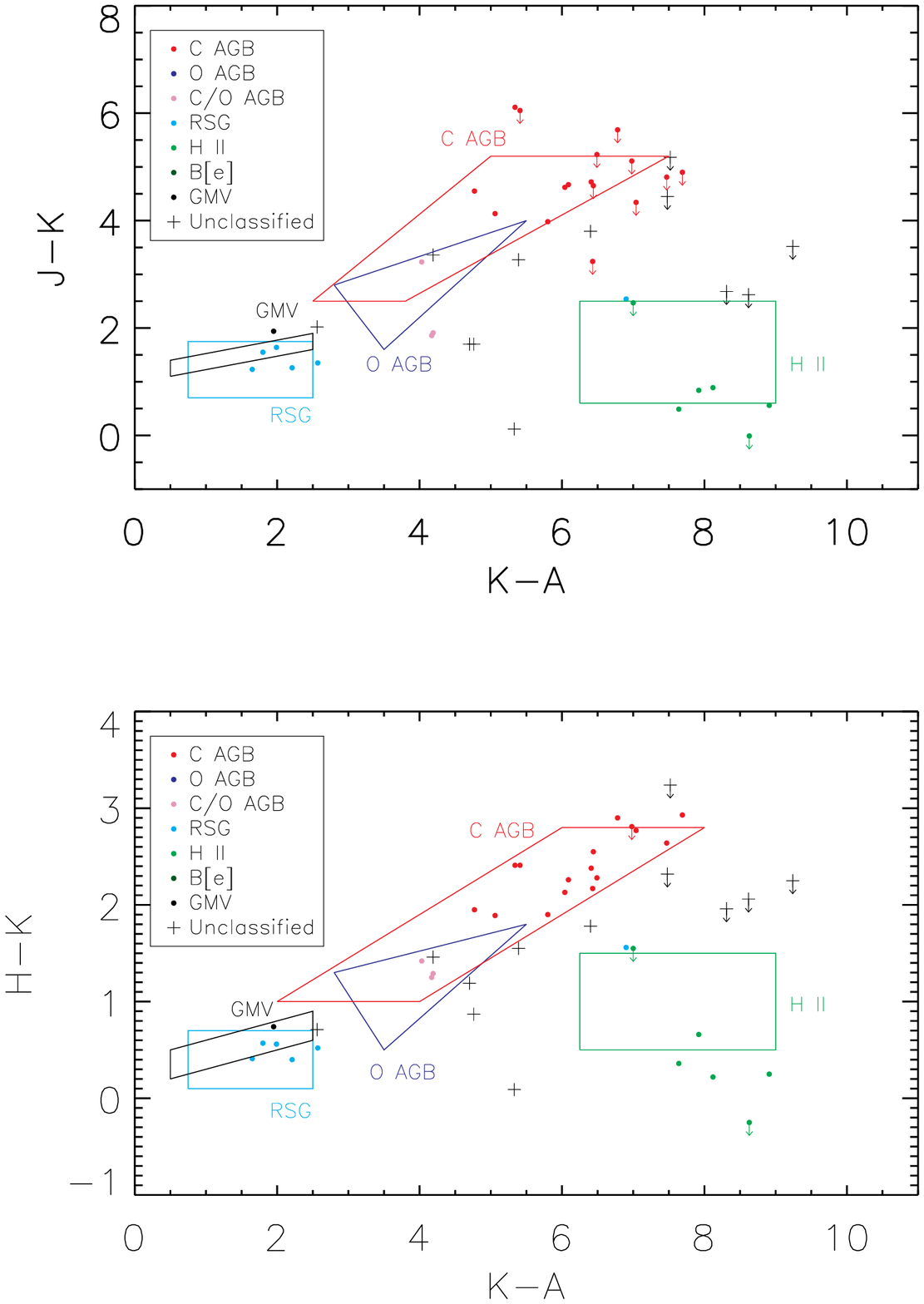}
\caption{JHK8 classifications of the available \protect\spitzer\ archival
  sources in the 2MASS/MSX color-color diagrams. Symbols indicate the
  JHK8 classification of each object from \protect\kas. Arrows indicate
  limits. The enclosed areas represent the classification regions
  defined in \protect\kas. \label{fig:colsam}} \epsscale{1.0}
\end{figure}

\begin{figure} 
\epsscale{0.7}
\plotone{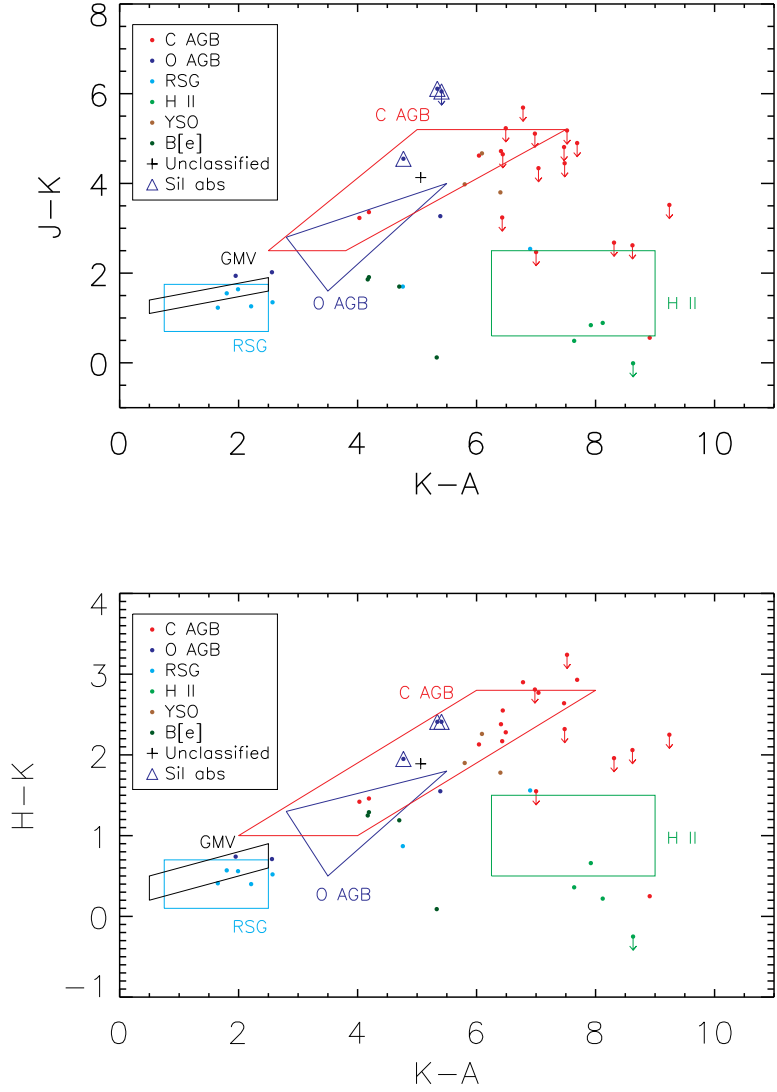}
\caption{2MASS/MSX color-color diagrams for the sources with confirmed
  spectroscopic classifications from archival \spitzer\ data. Colors
  indicate the class of each object (Table \ref{tab:class}). Arrows
  indicate limits and boxes indicate the JHK8 classification regions
  from \protect\kas. The three sources with silicate self-absorption,
  which have been classified as OH/IR stars by \citet{slo08}, are
  enclosed by triangles (see
  \protect\S\ref{subsubsec:dis_ohir}).\label{fig:colsp}}
\epsscale{1.0}
\end{figure}

We now compare the spectral classifications described in
\S\ref{sec:res} with the classifications of the same sources based on
JHK8 scheme (Paper II).  Where the JHK8 prediction (column 2 of Table
\ref{tab:class}) matches the actual classification (column 7), the
prediction is considered correct. JHK8 classifications are considered
incorrect where column 2 does not match column 7. For these purposes,
uncertain JHK8 predictions (indicated by a colon in column 2) are
considered the same as firm predictions.  Where the JKH8 class was
ambiguous due to overlaps in the classification boxes (e.g., RSG/GMV),
we consider the prediction to be ``correct but ambiguous'' if one of
the two classes in column 2 matches the classification in column 7.
Objects for which no JHK8 classification was possible (due to their
location outside the boxes in color-color plots) are considered
neither correct nor incorrect.  Figure \ref{fig:colsam} shows the
locations and JHK8 classifications of the 44 objects with published or
available archival spectra in the 2MASS/MSX color-color diagrams,
along with the JHK8 diagnostic regions used previously to classify the
object classes (\kas).  Figure \ref{fig:colsp} shows the IRS-based
spectral classifications of the archival objects in the 2MASS/MSX
color-color diagrams. 2MASS/MSX magnitudes of all sources are given in
\kas.

In summary, of the sample of 44 objects, the JHK8 scheme correctly
prediction the classification for 17 sources, and a further 5 had
correct but ambiguous classifications, where we have included C/O AGB
as an ambiguous match for the dusty early-type stars. Nine sources had
incorrect classifications. Twelve objects had no JHK8 classification
and, as noted earlier, one source could not be classified on the basis
of its IRS spectrum.

\begin{figure} 
\epsscale{0.7}
\plotone{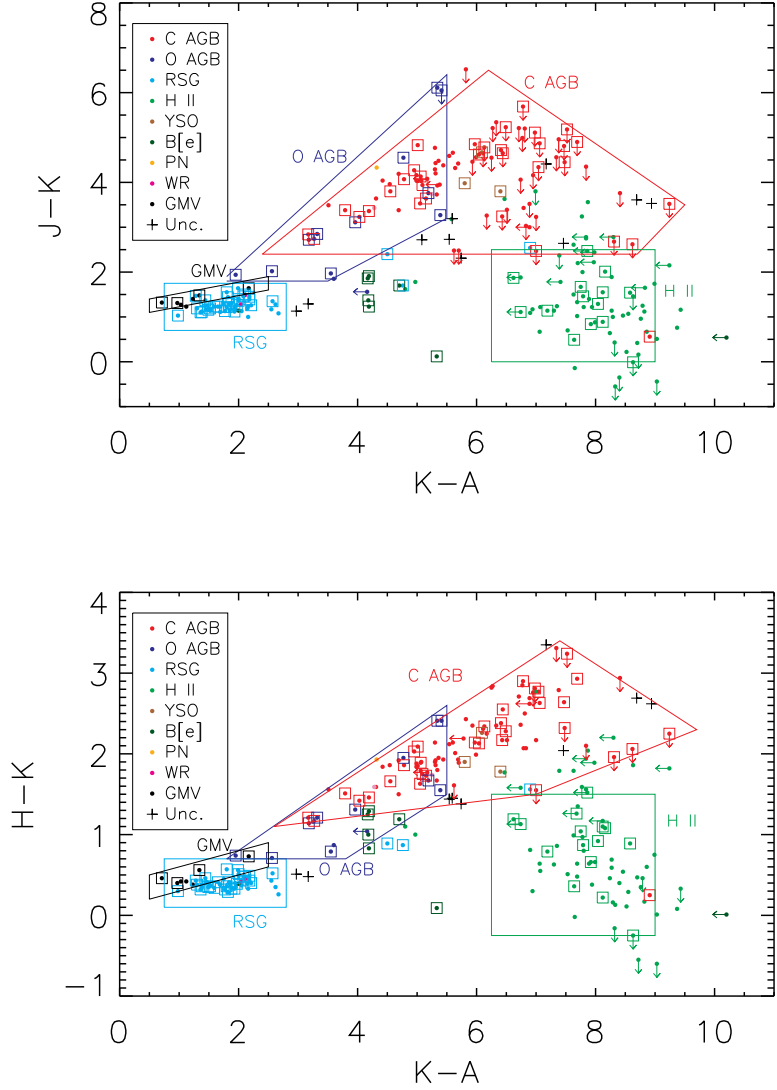}
\caption{2MASS/MSX color-color diagrams for the whole sample of the
  250 brightest 8~\micron\ sources in the LMC (\kas). Colors indicate
  the class of each object (from \protect\kas\ and Table
  \ref{tab:class}) and plus signs show objects which are
  unclassified. Arrows indicate limits. Open squares indicate objects
  whose classification is spectroscopically verified with IRS
  spectra. Boxes indicate the revised JHK8 classification
  regions.  \label{fig:colall}} \epsscale{1.0}
\end{figure}

In the following subsections we discuss the JHK8 classifications of
each object class in more detail.  We combine the new IRS-based
classification results with those obtained in Paper I so as to revise
and expand the JHK8 classification boxes.  The color-color diagrams of
the whole sample of 250 objects are shown in Figure \ref{fig:colall},
with the expanded classification boxes. The revised classification
criteria are listed in Table \ref{tab:crit}.

\begin{deluxetable}{lcc}
\rotate
\tabletypesize{\scriptsize}
\tablewidth{0pt}
\tablecolumns{3}
\tablecaption{Luminous LMC Mid-IR Sources: Revised JHK8 Color Classification Criteria \label{tab:crit}}
\tablehead{
\colhead{Class} &
\multicolumn{2}{c}{Criteria}
}
\startdata
\multicolumn{3}{c}{J$-$K vs. K$-$A colors}  \\
RSG              &  0.75 $\le$ (K$-$A) $\le$ 2.7                &  0.7 $\le$ (J$-$K) $\le$ 1.75  \\
O AGB            &  2.0 $\le$ (J$-$K); (K$-$A) $\le$ 5.5 &  [0.7 $\times$ (K$-$A) - 0.65] $\le$ (J$-$K) $\le$ [1.24 $\times$ (K$-$A) - 0.42]  \\
C AGB            & [0.93 $\times$ (J$-$K) + 0.16] $\le$ (K$-$A) $\le$ [0.73 $\times$ (J$-$K) + 6.95] &  2.4 $\le$ (J$-$K) $\le$ [$-$0.91 $\times$ (K$-$A) +12.15] \\
\protect\HII     &  6.25 $\le$ (K$-$A) $\le$ 9.0                &  0.0 $\le$ (J$-$K) $\le$ 2.5  \\
Expanded {\HII}  &  6.0 $\le$ (K$-$A) $\le$ 9.5                 &  (J$-$K) $\le$ 3.0  \\
GMV              &  0.5 $\le$ (K$-$A) $\le$ 2.5                 & [0.25 $\times$ (K$-$A) + 0.975] $\le$ (J$-$K) $\le$ [0.25 $\times$ (K$-$A) + 1.275] \\
\\
\multicolumn{3}{c}{H$-$K vs. K$-$A colors}  \\
RSG              &  0.75 $\le$ (K$-$A) $\le$ 2.7                &  0.1 $\le$ (H$-$K) $\le$ 0.7  \\
O AGB            &  0.7 $\le$ (H$-$K); (K$-$A) $\le$ 5.5 &  [0.47 $\times$ (K$-$A) - 1.09] $\le$ (H$-$K) $\le$ [0.51 $\times$ (K$-$A) - 0.21]  \\
C AGB            & [2.08 $\times$ (H$-$K) + 0.31] $\le$ (K$-$A) $\le$ [3.33 $\times$ (H$-$K) + 2.0] &  [0.09 $\times$ (K$-$A) + 0.86] $\le$ (H$-$K) $\le$ [$-$0.48 $\times$ (K$-$A) +6.94] \\
\protect\HII     &  6.25 $\le$ (K$-$A) $\le$ 9.0                &  $-$0.25 $\le$ (H$-$K) $\le$ 1.5  \\ 
Expanded {\HII}  &  6.0 $\le$ (K$-$A) $\le$ 9.5                 &  (H$-$K) $\le$ 2.0   \\
GMV              &  0.5 $\le$ (K$-$A) $\le$ 2.5                 &  [0.2 $\times$ (K$-$A) + 0.1] $\le$ (H$-$K) $\le$ [0.2 $\times$ (K$-$A) + 0.4]  \\
\enddata 
\end{deluxetable}

\begin{deluxetable}{lcc}
\rotate
\tabletypesize{\scriptsize}
\tablewidth{0pt}
\tablecolumns{3}
\tablecaption{Luminous LMC Mid-IR Sources: Revised IRAC/MIPS Color Classification Criteria \label{tab:critsage}}
\tablehead{
\colhead{Class} &
\multicolumn{2}{c}{Criteria}
}
\startdata
\multicolumn{3}{c}{[3.6]$-$[4.5] vs. [5.8]$-$[8.0] colors}  \\
RSG              &  [-1.00  $\times$ ([3.6]$-$[4.5]) + 0.20] $\le$ [5.8]$-$[8.0] $\le$ [-1.00 $\times$ ([3.6]$-$[4.5]) + 1.00] & [1.29 $\times$ ([5.8]$-$[8.0]) - 0.95] $\le$ [3.6]$-$[4.5] $\le$[1.29 $\times$ ([5.8]$-$[8.0]) - 0.37] \\
O AGB            &  [-1.00  $\times$ ([3.6]$-$[4.5]) + 1.40] $\le$ [5.8]$-$[8.0] $\le$ [-1.25 $\times$ ([3.6]$-$[4.5]) + 2.06] & [9.00 $\times$ ([5.8]$-$[8.0]) - 10.6] $\le$ [3.6]$-$[4.5] $\le$[1.25 $\times$ ([5.8]$-$[8.0]) - 0.40] \\
C AGB            &  [-0.50  $\times$ ([3.6]$-$[4.5]) + 0.73] $\le$ [5.8]$-$[8.0] $\le$ [-0.43 $\times$ ([3.6]$-$[4.5]) + 2.14] & [1.39 $\times$ ([5.8]$-$[8.0]) - 0.42] $\le$ [3.6]$-$[4.5] $\le$[1.44 $\times$ ([5.8]$-$[8.0]) - 0.08] \\
\\
\multicolumn{3}{c}{[5.8]$-$[8.0] vs. [8.0]$-$[24] colors}  \\
RSG              &  [-2.12 $\times$ ([5.8]$-$[8.0]) + 2.33] $\le$ [8.0]$-$[24] $\le$ [-2.67 $\times$ ([5.8]$-$[8.0]) + 4.59] & [0.38 $\times$ ([8.0]$-$[24]) -0.43] $\le$ [5.8]$-$[8.0] $\le$[0.55 $\times$ ([8.0]$-$[24]) -0.18] \\
O AGB            &  [-5.94 $\times$ ([5.8]$-$[8.0]) + 8.15] $\le$ [8.0]$-$[24] $\le$ [-1.11 $\times$ ([5.8]$-$[8.0]) + 4.14] & [5.8]$-$[8.0] $\le$ [0.53 $\times$ ([8.0]$-$[24]) -0.13] \\
C AGB            & 0.42 $\le$ ([5.8]$-$[8.0]) ; [8.0]$-$[24] $\le$ [-0.45 $\times$ ([5.8]$-$[8.0]) + 3.04] & [0.56 $\times$ ([8.0]$-$[24]) -0.20] $\le$ [5.8]$-$[8.0] $\le$ [0.41 $\times$ ([8.0]$-$[24]) +0.44] \\
\\
\multicolumn{3}{c}{J$-$K vs. [3.6]$-$[4.5] colors}  \\
RSG              &  [-0.08  $\times$ (J$-$K) - 0.09] $\le$ [3.6]$-$[4.5] $\le$ [-0.06 $\times$ (J$-$K) + 0.41] & [0.40 $\times$ ([3.6]$-$[4.5]) + 0.86] $\le$ J$-$K $\le$[0.39 $\times$ ([3.6]$-$[4.5]) + 1.68] \\
O AGB            &  [0.10  $\times$ (J$-$K) - 0.05] $\le$ [3.6]$-$[4.5] $\le$ [0.15 $\times$ (J$-$K) + 0.17] & 1.9 $\le$ J$-$K $\le$ 6.2 \\
C AGB           &  [0.50  $\times$ (J$-$K)] $\le$ [3.6]$-$[4.5] $\le$ [0.18 $\times$ (J$-$K) - 0.05] & [-0.75 $\times$ ([3.6]$-$[4.5]) + 3.02] $\le$ J$-$K $\le$[4.27 $\times$ ([3.6]$-$[4.5]) + 12.54] \\
\enddata
\end{deluxetable}

\subsection{Oxygen-rich objects} \label{subsec:dis_orich}

Of the 17 objects with O-rich dust chemistry, 12 had JHK8
classifications (Table \ref{tab:class}). Of these, 3 classifications
were correct and a further 4 were ambiguously correct, while 5 were
incorrect.  The classes of O-rich object are discussed in the
following subsubsections.

\subsubsection{O-rich asymptotic giant branch stars}

Of the three sources classified as O-rich AGB stars on the basis of
IRS spectra (Table \ref{tab:class}), MSX\,LMC~1192 had a tentative
JHK8 classification of GMV, while MSX\,LMC~283 and 1190 were
previously unclassified by the JHK8 scheme. Thus none of these three
sources lies in the previous (\kas) O AGB classification box.

The IR colors of MSX\,LMC~283 (see Table 1 of \kas) place it redward
(in K-A) of the ambigous overlap region of the O-rich and C-rich AGB
stars. On the basis of the location of MSX\,LMC~283, we extend the O
AGB box to the right in K-A (Figure \ref{fig:colall}).  As the overlap
area between O-rich and C-rich AGB stars now covers a large fraction
of the region occupied by O-rich AGB stars, the classification of the
majority of the O-rich AGB stars in a IR-luminous stellar population
solely on the basis of their JHK8 colors will be ambiguous. However,
the increased overlap is largely due to the location of the OH/IR
stars (\S\ref{subsubsec:dis_ohir}) and many of the O-rich AGB stars
lie outside the overlap region. In addition, as discussed in \kas, the
O-rich and C-rich stars seem to be well-distinguished by their
[8]-[24]~\micron\ SAGE colors (see \S\ref{subsec:dis_spitzer}).

MSX\,LMX~1190 and 1192 lie near the RSG classification box in the JHK8
scheme, bluer in K-A than the O-rich AGB box by $\sim$1 mag.  On the
basis of the colors of MSX\,LMC~1190 and 1192, we extend the O AGB
box to bluer K-A, covering the region between the previous O-rich AGB
star and RSG classification boxes, and leading to a small overlap
between the O AGB and GMV boxes. For objects in this region,
additional information, such as the K magnitude, is necessary to
determine the nature of the object.

\subsubsection{OH/IR stars} \label{subsubsec:dis_ohir}

The three OH/IR stars, MSX\,LMC~811, 936, and 1171, were all
incorrectly classified by the JHK8 scheme as carbon-rich AGB stars.
These stars show silicate self-absorption indicative of
optically-thick dust shells due to high mass-loss rates.  Their
locations in the color-color diagrams (Figure \ref{fig:colsp}) are
indicative of increasingly strong self-absorption due to high
mass-loss rates.

These objects lie above most carbon stars in the 2MASS/MSX color-color
diagrams (i.e., for a given K-A color, the three OH/IR stars are
generally redder in J-K and H-K than carbon stars) but are
nevertheless within or above the JHK8 diagnostic box for C-rich AGB
stars, suggesting that for modestly self-absorbed objects,
spectroscopy or additional photometry may be necessary to distinguish
OH/IR stars from carbon stars.  The O-rich AGB classification box has
been expanded to include these objects, increasing the ambiguous
overlap region between carbon-rich and oxygen-rich objects
(Fig. \ref{fig:colall}).

\subsubsection{Red Supergiants}

Of the 7 RSGs, three were correctly classified as RSGs on the basis of
JHK8 colors and two (MSX\,LMC~461 and 1117) had correct but ambiguous
classifications of RSG/GMV.  One object, MSX\,LMC~500, was incorrectly
classified as an \htwo\ region and one object, MSX\,LMC~886, was not
classifiable via JHK8 colors.

MSX\,LMC~500 was most likely misclassified due to the surrounding
nebulous dust emission contaminating the JHK8 colors (see
\S\ref{subsubsec:res_rsg}), as it lies close to the \htwo\ region
classification box, $\sim$5 magnitudes redder in K-A than typical
RSGs.  MSX\,LMC~886 lies $\sim$2 magnitudes redder in K-A than typical
RSGs, quite close to the O-rich AGB classification box.  The \kas\ RSG
classification box covers the region of color-color space occupied by
most of the RSGs and was not expanded or altered on the basis of the
position of MSX\,LMC~500 or 886.

\subsubsection{Dusty, early-type stars}

Of the four dusty, early-type stars, two had correct but ambiguous
JHK8 classifications, and two were unclassifiable via the JHK8
scheme. The former two, MSX\,LMC~262 and 887, lie within the O-rich
AGB and C-rich AGB classification boxes in H-K vs. K-A but not J-K vs
K-A (Fig.\ \ref{fig:colall}).  The latter two objects, MSX\,LMC~323
and 134, have similar colors to 262 and 887 but lie outside the
classification boxes. No dusty, early-type star classification box is
defined, due to the small number of objects.

\subsection{Carbon-rich AGB stars} \label{subsec:dis_cagb}

Of the 19 C-rich sources, 10 were correctly classified as carbon-rich
on the basis of their JHK8 colors, and one (MSX\,LMC~1130) was
correctly but ambiguously classified as either C- or O-rich.  Six
objects lay outside the classification boxes in Figure
\ref{fig:colsam} and so had no JHK8 classification, while two sources
(MSX\,LMC~219 and 1302) were incorrectly classified as \htwo\ regions
on the basis of their JHK8 colors.

The 6 C-rich AGB stars which could not be classified under the JHK8
system lie near the C-rich AGB diagnostic box (Figure \ref{fig:colsp})
but have redder JKH8 colors.  This region of color space was not
covered by our \buc\ observations but, on the basis of the new
spectral data, we may now extend the C-rich AGB star classification
region to include objects with K-A$\gtrsim$6.5, J-K$\gtrsim$2.5, and
H-K$\gtrsim$1.5. The revised C AGB box covers the space between the
\kas\ C AGB box and the \htwo\ region classification box and results
in a small overlap between the boxes (Figure \ref{fig:colall}).  For
objects with colors in the overlap region, imaging aimed at
distinguishing between diffuse vs.\ point-like emission would be
essential in the absence of spectroscopy to determine unambiguous
classifications.

One of the two C-rich AGB stars which was incorrectly classified as an
\htwo\ region (MSX\,LMX~219) lies in the overlap region between the C
AGB and \htwo\ region boxes. The other star, MSX\,LMC~1302, lies well
within the \htwo\ region classification box and away from the other
C-rich AGB stars. The C AGB box was not extended to include 1302, due
to the uncertainty of its classification and, in particular, the
possibility that it may be a pre-planetary nebula.

\subsection{Compact \protect\htwo\ regions} \label{subsec:dis_hii}

For the four confidently classified \htwo\ regions, the JHK8 color
scheme correctly predicted the spectral classification. The location
of these objects expands the \htwo\ region classification box to bluer
J-K and H-K colors (Fig. \ref{fig:colall}).  

\subsection{Young Stellar Objects and Candidates} \label{subsec:dis_yso}

Of the three YSO or YSO candidates in the sample, two were incorrectly
classified as C-rich AGB stars by the JHK8 scheme and the other had no
JHK8 classification.  Although they resemble \htwo\ regions in their
mid-IR spectra, MSX\,LMC~46, 771, and 1184 have redder J-K and H-K
colors than \htwo\ regions.  All three objects lie near each other and
within the revised C AGB classification box.  No YSO classification
box is defined due to the small number of objects and the uncertain
nature of their classifications.

\subsection{\protect\spitzer\ IRAC/MIPS classification of IR-luminous
  point sources} \label{subsec:dis_spitzer}

\begin{figure} 
\epsscale{0.7}
\plotone{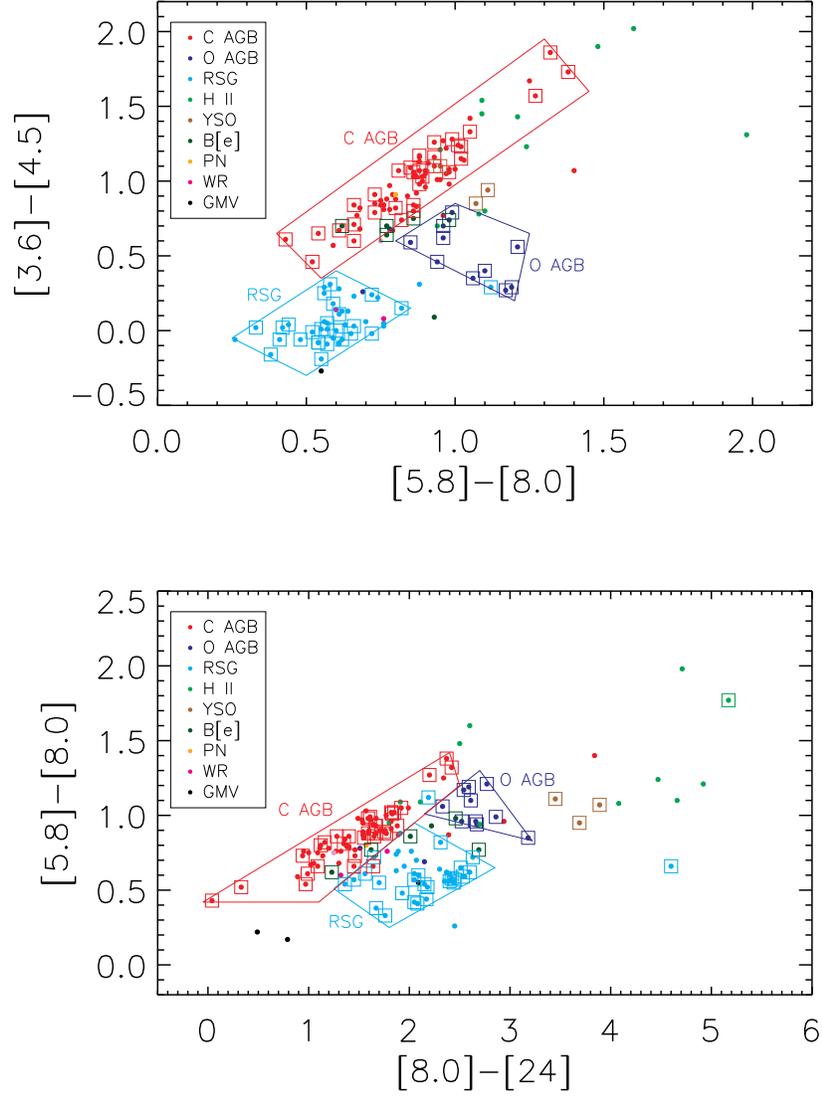}
\caption{\protect\spitzer\ IRAC/MIPS color-color diagrams for the
  whole sample of the 250 brightest 8~\micron\ sources in the
  LMC. Colors indicate the class of each object (from
  \protect\kas\ and Table \ref{tab:class}). Open squares indicate
  objects whose classification is spectroscopically verified with IRS
  spectra. Boxes indicate the revised classification
  regions.  \label{fig:sagecolall}} \epsscale{1.0}
\end{figure}

\begin{figure} 
\epsscale{0.7}
\plotone{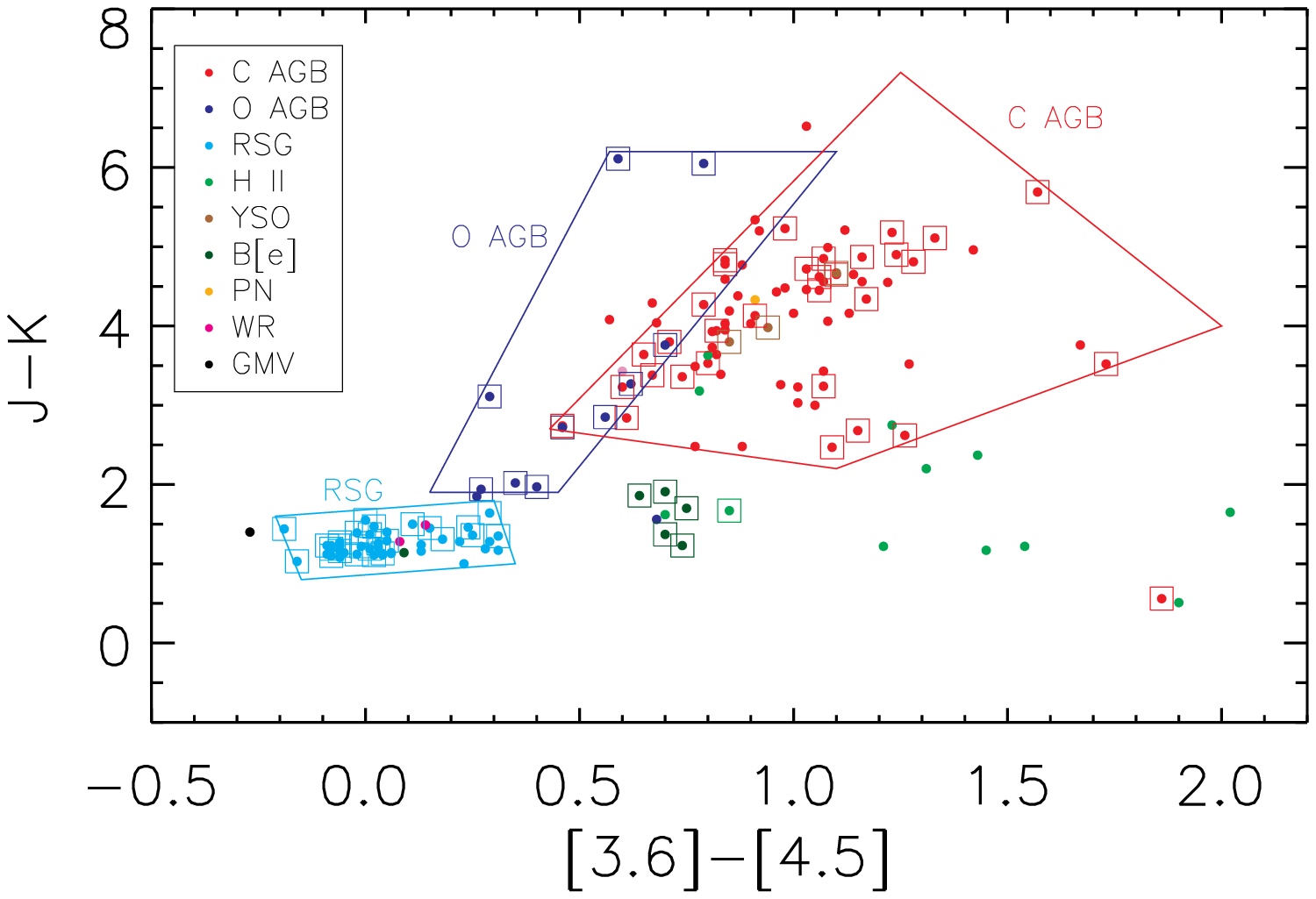}
\caption{\protect\spitzer\ IRAC and 2MASS J-K color-color diagram for
  the whole sample of the 250 brightest 8~\micron\ sources in the
  LMC. Colors indicate the class of each object (from
  \protect\kas\ and Table \ref{tab:class}). Open squares indicate
  objects whose classification is spectroscopically verified with IRS
  spectra. Boxes indicate classification regions defined from the
  spectroscopically verified objects. \label{fig:spitzerwarm}}
\epsscale{1.0}
\end{figure}

Figure \ref{fig:spitzerwarm} shows \spitzer\ IRAC/MIPS color-color
diagrams for the sources in the whole sample of 250 objects with SAGE
counterparts (see \kas). Diagnostic regions defined by the
spectroscopically classified objects are shown in Figure
\ref{fig:spitzerwarm} and the corresponding IRAC/MIPS classification
criteria for luminous IR sources, revised according to the IRS
spectral classification results reported here, are listed in Table
\ref{tab:critsage}. As noted in \kas, \spitzer\ IRAC and IRAC/MIPS
color-color diagrams such as those in Fig. \ref{fig:sagecolall}
provide better discrimination between carbon-rich and oxygen-rich AGB
stars than do JHK8 color-color diagrams. This discrimination has been
further improved by the IRS spectral classifications, which resulted
in the reclassification of several apparent C-rich AGB stars as OH/IR
stars (Table \ref{tab:class}). Figure \ref{fig:spitzerwarm} shows the
\spitzer\ IRAC [3.6]-[4.5] color plotted against 2MASS J-K. While this
diagram shows somewhat more overlap between the C-rich and O-rich AGB
stars than do the \spitzer\ IRAC and IRAC/MIPS diagrams
(Fig. \ref{fig:sagecolall}), the J-K vs.  [3.6]-[4.5] diagram will
remain useful even after Spitzer enters its warm mission, in which
only the 3.6 and 4.5~\micron\ channels will operate. We note that the
outlying C-rich AGB star in the bottom right corner of the diagram
(MSX\,LMC~1302) is excluded from the `C AGB' region as its colors may
be affected by extended emission. Finally, we note that the two O-rich
sources with J-K$\sim$6 in this diagram are OH/IR stars that show
silicate self-absorption.

\section{SUMMARY AND CONCLUSIONS} \label{sec:con}

We have used archival and published IRS spectra to classify 43
IR-luminous sources in the LMC. Of the 31 objects of these 43 that
have a previous (\kas) JHK8 classification, we find that the JHK8
classifications were correct for 22 objects (71\%).  Of the nine
objects with incorrect JHK8 classifications, four had tentative
classifications.

Spectroscopic classifications of the 12 objects which were previously
unclassifiable with the JHK8 diagnostics allow us to characterize
new regions of the 2MASS/MSX color-color diagrams. In addition,
analysis of the IRS spectra obtained for the 9 objects which were
previously incorrectly classified under the JHK8 scheme allows us to
revise the classification boxes so as to make JHK8 color-based
classifications more reliable. The area of color space occupied by
C-rich AGB stars is extended to redder K-A colors, towards the
location of \htwo\ regions. Similarly --- as a consequence of the
reclassification of 3 objects from C-rich AGB (Paper II) to OH/IR
stars (Sloan et al.\ 2008) --- the region of color space occupied by
O-rich stars has been extended, increasing the overlap area between
carbon-rich and oxygen-rich stars. The \htwo\ region area has also
been expanded, to bluer J-K and K-A colors. The RSG region has been
extended to redder K-A colors.

Those sources whose classifications were incorrect under the Paper II
JHK8 system offer insight into the use of color-based diagnostics to
identify the nature of objects. Many of these objects lie in overlap
regions between diagnostic boxes or towards the edges of boxes. In
particular, the revised JHK8 scheme presented here reveals larger
overlap in color space between AGB stars of different dust chemistries
(C-rich vs.\ O-rich), due to the inclusion of a handful of previously
misclassified OH/IR stars. In the LMC, this overlap region is
dominated by carbon stars (Fig.\ \ref{fig:colall}) 
as a consequence of the LMC's low metallicity (Paper II), resulting in
a relatively low JHK8 misclassification rate overall
($\sim$20\%). However, in the Milky Way and external galaxies of
similar metallicity, more caution must be exercised in identifying the
nature of sources as C-rich vs.\ O-rich AGB stars based on their JHKA
colors alone. Furthermore, ``contamination'' of photometry or spectra
due to crowded fields or surrounding \htwo\ regions confuses the
classification of sources located in or near star-forming regions.

However, with the important exception of O-rich AGB stars with high
mass loss rates (OH/IR stars) -- which are likely to be present in
larger proportion (relative to high mass loss rate carbon stars) in
external galaxies with higher metallicity than that of the LMC -- the
results presented here indicate that the revised JHK8 color
classification criteria can be used to classify the most luminous IR
sources in nearby galaxies with $>70$\% confidence.  These results
therefore reinforce both the utility of the JHK8 color diagnostics and
the conclusions in Paper II as to the preponderance of C-rich AGB
stars and \htwo\ regions among most luminous IR sources in the LMC.
In addition, the spectroscopic identifications found in this paper
further reinforce the results of \kas\ that \spitzer\ IRAC/MIPS
color-color diagrams and IRAC/2MASS color-color diagrams can be used
to distinguish between object classes. In particular, color-color
diagrams that make use of Spitzer photometry appear to be far more
effective in discriminating between O-rich and C-rich AGB stars than
diagrams based solely on JHK8 colors.

\acknowledgments

This work is based on data from the \emph{Spitzer} Space Telescope,
which is operated by the Jet Propulsion Laboratory, California
Institute of Technology under a contract with NASA.  This research has
made use of the SIMBAD database, operated at CDS, Strasbourg, France.
We thank the anonymous referee for considered comments which improved
this manuscript.

{\it Facilities:} \facility{\spitzer}

\end{document}